\documentclass[prl,reprint,superscriptaddress,longbibliography]{revtex4-2}

\usepackage{amsmath}
\usepackage{amssymb}
\usepackage{graphicx}
\usepackage{hyperref}
\usepackage{bm}

\hypersetup{pdftitle={Transport Signatures of a Kinetically Self-Trapped Nagaoka Polaron},
            pdfauthor={Yusong Xiao and Yuan Wan}}
\begin{document}
\title{Transport Signatures of a Kinetically Self-Trapped Nagaoka Polaron}

\date{September 29, 2026}

\author{Yusong Xiao}
\affiliation{Institute of Physics, Chinese Academy of Sciences, Beijing 100190, China}
\affiliation{University of Chinese Academy of Sciences, Beijing 100049, China}

\author{Yuan Wan}
\email{yuan.wan@iphy.ac.cn}
\affiliation{Institute of Physics, Chinese Academy of Sciences, Beijing 100190, China}
\affiliation{Songshan Lake Materials Laboratory, Dongguan, Guangdong 523808, China}

\begin{abstract}
Nagaoka polarons are canonical examples of kinetic magnetism. Although their spin structure is well understood, their transport remains unexplored. We compute the dynamical mobility of a single hole in the square-lattice $t$-model using resummation-based quantum Monte Carlo and analytic continuation. Upon cooling, an incoherent Drude-like response gives way to a finite-frequency peak as the dc mobility vanishes. The peak signals an internal mode of a hole confined by its ferromagnetic cloud. Our results establish transport as an incisive probe of polaron formation and kinetic self-trapping.
\end{abstract}

\maketitle

\begin{figure}
\includegraphics[width = \columnwidth]{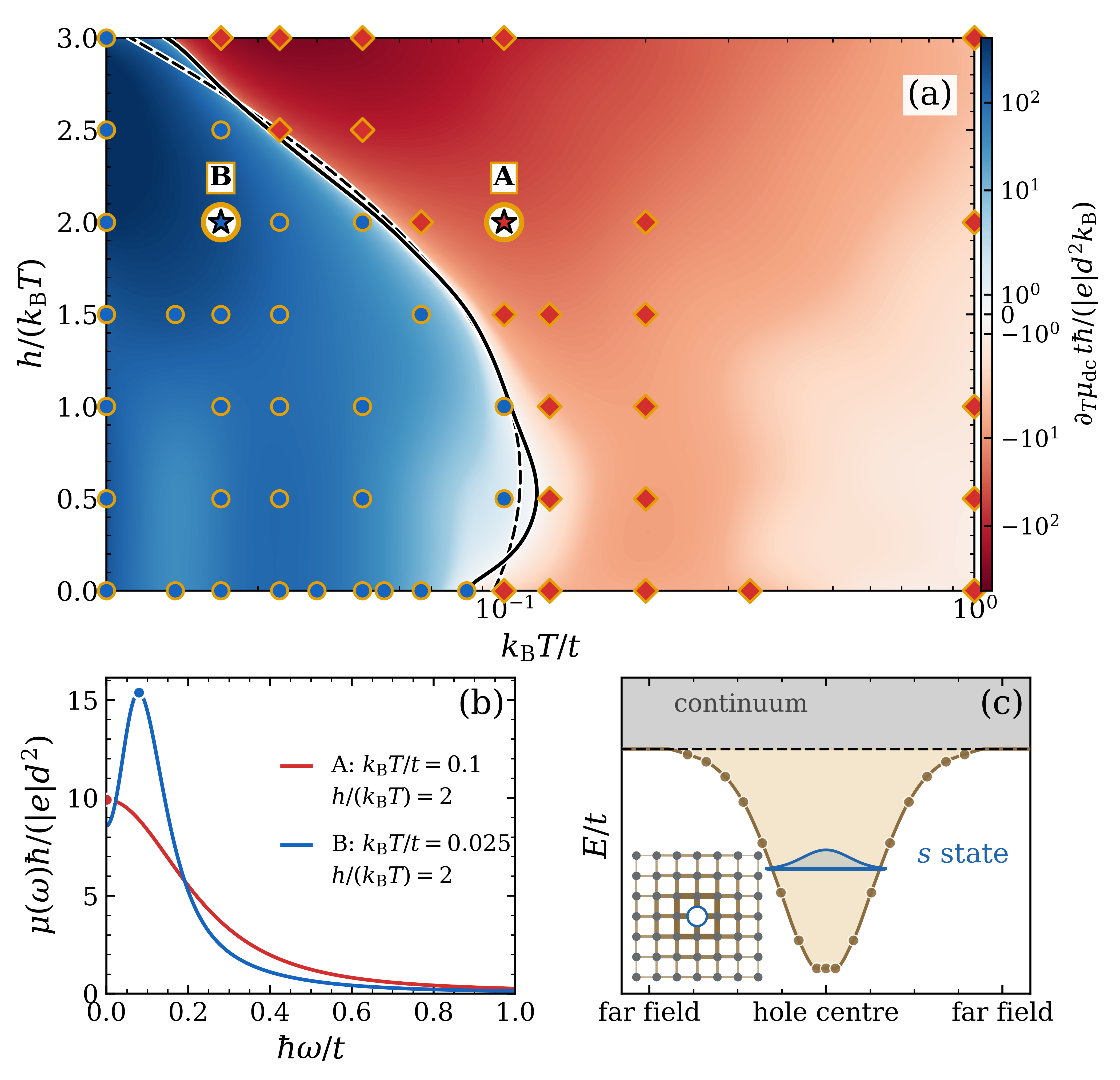}
\caption{Summary of the main results. (a) The temperature derivative of the DC mobility $\partial\mu_{DC}/\partial T$ as a function of temperature $T$ and Zeeman field $h$. Blue and red shaded areas are the polaron ($\partial\mu_{DC}/\partial T > 0$) and the incoherent transport regimes ($\partial\mu_{DC}/\partial T < 0$), separated by a crossover (solid line). The dashed line shows the crossover determined by the appearance of a finite frequency peak in the dynamical mobility. (b) Representative dynamical mobility in the polaron regime (blue) and the incoherent transport regime (red). The corresponding locations in (a) are marked. (c) The finite frequency dynamical mobility peak is attributed to the internal modes of the hole trapped in a soft attraction potential.}
\label{fig:phase}
\end{figure}

\textit{Introduction ---} Kinetic magnetism is a quantum phenomenon where the spin correlations arise in a magnet not by exchange but by the motion of doped charge~\cite{Nagaoka1966,Tasaki1989,HaerterShastry2005}. The cleanest example is a Nagaoka polaron~\cite{Nagaoka1966,Tasaki1989,Powell2017}, where the doped charge lowers its kinetic energy by polarizing the surrounding spins, dressing itself in a ferromagnetic spin cloud. Once a purely theoretical concept, kinetic magnetism has recently been observed in quantum simulators~\cite{Dehollain2020,Xu2023,QiaoRydberg2025,Martin2026} and moir\'{e} systems~\cite{Tang2020,Ciorciaro2023,Tao2024,Xiong2024}, allowing for direct imaging of the Nagaoka polaron~\cite{Lebrat2024,Prichard2024}.

To date, the Nagaoka polaron~\cite{Dagotto1989,WhiteAffleck2001,Yun2023,SamajdarBhatt2024,BotzungNataf2024,NewbyKhatami2025} and its antiferromagnetic cousins~\cite{HaerterShastry2005,WangGongWang2008,Sposetti2014,Lisandrini2017,ZhangBatista2018,Blesio2019,Morera2023,Kim2023,PereiraMueller2025,Carlstrom2025,Pei2026,PereiraMueller2026,Sharma2026} have been almost exclusively characterized by their spin structures. As the temperature decreases, the spin correlations build up gradually in magnitude and range near the charge. Yet, their smooth thermal evolution yields no clear sign for the formation of the Nagaoka polaron. Meanwhile, little is known about its transport properties even though transport is the defining feature of a polaron. Early studies established the linear-in-temperature resistivity based on the picture that the charge avoids scrambling the spins by retracing its previous steps~\cite{BrinkmanRice1970,OguriMaekawa1990,MetznerSchmitVollhardt1992}. This picture breaks down at lower temperature when the polaron forms due to scrambling. Exact diagonalization studies~\cite{JaklicPrelovsek1995} are limited by finite size effects. More recent works explore the nonequilibrium dynamics~\cite{Carlstrom2016,KanaszNagy2017} or transport at finite doping~\cite{Perepelitsky2016,Cao2026,Tulipman2026}. The transport of a Nagaoka polaron, despite its simple setup and conceptual significance, has surprisingly escaped theoretical scrutiny.

In this work, we uncover the transport of a Nagaoka polaron through the lens of dynamical mobility. We consider the canonical setting of a hole in the square lattice $t$-model~\cite{Nagaoka1966,Tasaki1989}. When the system cools, the dynamical mobility transforms from an incoherent Drude-like peak to an insulating response with a finite frequency peak (Fig.~\ref{fig:phase}(b)). We identify this peak as the internal mode of the hole moving in the soft attraction potential (Fig.~\ref{fig:phase}(c)): As the hole lowers its kinetic energy by aligning neighboring spins, its motion is also weakly confined by the spin cloud. Therefore, in sharp contrast to the monotonically growing spin correlations, the dynamical mobility provides unambiguous signatures that mark the formation of the Nagaoka polaron (Fig.~\ref{fig:phase}(a)), establishing the transport as an incisive probe of kinetic self-trapping.

We construct the dynamical mobility of the Nagaoka polaron from quantum Monte Carlo (QMC) data through stochastic analytic continuation (SAC)~\cite{Sandvik1998,Shao2023,Sandvik2016}. We extend an efficient resummation-based QMC method~\cite{MullerHartmann1994,RaghavanElser1995,PereiraMueller2025} such that it delivers high quality current correlation function data, thus making the continuation robust. We validate the SAC results by showing that the spectral features are fully consistent with a continuation-free spectral reconstruction method~\cite{HLT2019}. 

\textit{Setup ---} We consider the square lattice $t$-model in the single hole sector:
\begingroup
\medmuskip=3mu
\thickmuskip=3mu
\begin{align}
H = -t\sum_{ij \sigma} \mathcal{P} c^\dagger_{i\sigma} c^{\phantom\dagger}_{j\sigma} \mathcal{P}- h \sum_i S^z_i/\hbar \equiv -t H_t - hM^z.
\end{align}
\endgroup
$t>0$ is the hopping amplitude on the directed, nearest neighbor bond $ij$. $c_{i\sigma}$ ($c^\dagger_{i\sigma}$) annihilates (creates) an electron of spin $\sigma$ on lattice site $i$. $\mathcal{P}$ projects out the doubly occupied states. We use the Zeeman field $h$ as a controlling knob. $S^z_i = \hbar (c^\dagger_{i\uparrow}c^{\phantom\dagger}_{i\uparrow}-c^\dagger_{i\downarrow}c^{\phantom\dagger}_{i\downarrow})/2$ is the local magnetic moment in the $z$ direction. $[H_t,M^z] = 0$.

The path integral representation of the partition function $Z = \mathrm{Tr}e^{\beta t H_t + \beta h M^z}$ is given by: 
\begin{align}
\frac{Z}{\Omega} = \sum_{n}\frac{(\beta t)^n}{n!}\frac{\mathrm{Tr}(e^{\beta h M^z} H^n_t)}{\Omega}  = \sum_{\Gamma} \frac{(\beta t)^{n_\Gamma}}{n_\Gamma!} f(\Gamma).
\label{eq:Z}
\end{align}
We rescale $Z$ by the partition function of a freely placed hole and $V-1$ independent spins, $\Omega = Vz^{V-1}(\beta h)$, with $V$ being the number of sites. $z(\beta h) = 2\cosh(\beta h/2)$ is the partition function of a spin. The second equality follows from the fact that the powers of $H_t$ generate closed walks $\Gamma$ in the square lattice. $n_\Gamma$ is its length. The initial/final position of $\Gamma$ is set to the origin by translation symmetry. 

The spin factor $f(\Gamma)$ is defined as follows~\cite{RaghavanElser1995,PereiraMueller2025}: As the hole exchanges its position with a spin in each step, a closed walk $\Gamma$ induces a spin permutation $\Pi_\Gamma$:
\begin{align}
f(\Gamma) \equiv \frac{\mathrm{Tr}'(e^{\beta h M^z}\Pi_\Gamma)}{z(\beta h)^{V-1}} = \frac{\prod^C_{\alpha=1}z(l_\alpha \beta h)}{z(\beta h)^{V-1}}.
\label{eq:f_weight}
\end{align}
The trace is over $2^{V-1}$ spin states with the hole fixed at the origin. We compute $f(\Gamma)$ by decomposing $\Pi_\Gamma$ into $C$ independent cycles, labeled by $\alpha=1,2,\cdots, C$. Let $l_\alpha$ be the length of cycle $\alpha$. The spins in the same cycle $\alpha$ must align ferromagnetically; otherwise, such states do not contribute to the trace. Therefore, they form a block spin with magnetic moment $l_\alpha/2$, whose contribution to the trace is $z(l_\alpha \beta h)$. Collecting all cycles produces the second expression.

\begin{figure}
\includegraphics[width = \columnwidth]{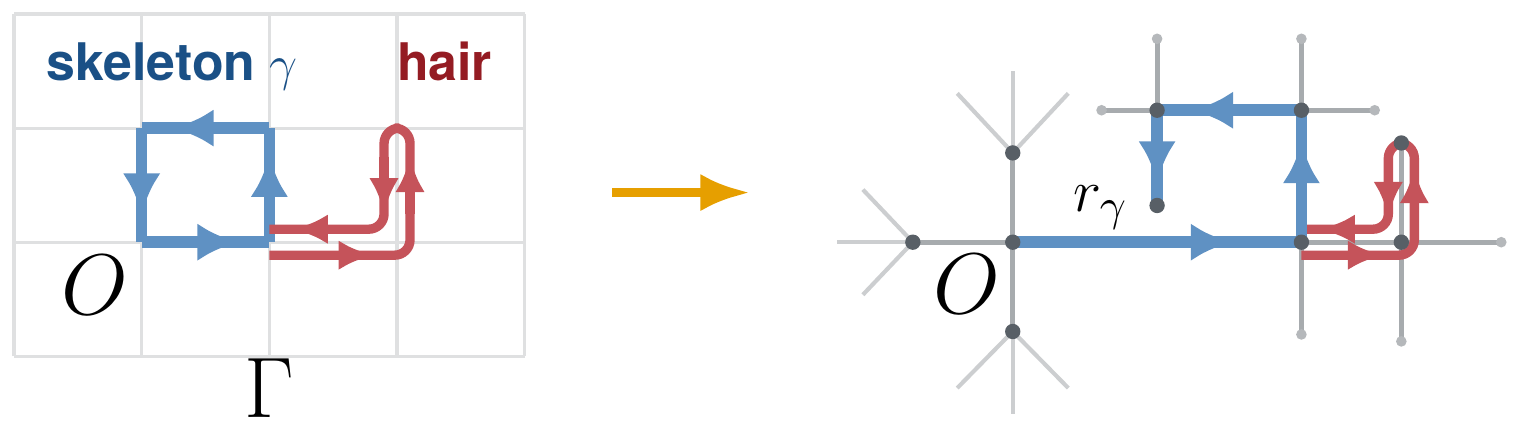}
\caption{A walk $\Gamma$ on the square lattice is decomposed into a skeleton (blue) dressed by self-retracing hairs (red). When embedded in the Bethe lattice, the skeleton is mapped to a Bethe lattice site, whereas the hair is mapped to excursions.}
\label{fig:bethe_lattice}
\end{figure}

\textit{Resummation ---} The path integral Eq.~\eqref{eq:Z} is made more efficient by a partial summation~\cite{MullerHartmann1994,RaghavanElser1995}. We write a walk $\Gamma$ as a string of letters, $\Gamma = s_1s_2\cdots $. $s_i\in \{x,\overline{x},y,\overline{y}\}$, representing the four directions in each step. A self-retracing walk is of the form $s_1s_2\cdots s_k \overline{s}_k\cdots \overline{s}_2\overline{s}_1$, where $\overline{s}$ denotes reversal. Such walks do not permute the spins, and, therefore, dressing any given walk $\Gamma$ by them does not change its weight $f(\Gamma)$. We thus rewrite Eq.~\eqref{eq:Z} as:
\begin{align}
\frac{Z}{\Omega} = \sum_{\gamma}f(\gamma) (\sum_{\Gamma \uparrow \gamma}  \frac{(\beta t)^{n_\Gamma}}{n_\Gamma!}) \equiv \sum_{\gamma}f(\gamma) w(\gamma).
\label{eq:Z_resum}
\end{align}
The outer summation is over all closed walks $\gamma$ with no immediate reversals, known as the skeletons. The inner sum is over all possible walks $\Gamma$ obtained by dressing $\gamma$ with self-retracing segments, known as hairs (Fig.~\ref{fig:bethe_lattice}). The second equality defines the kinetic weight of the skeleton $\gamma$.

Previous works use combinatorics to find the kinetic weight $w(\gamma)$~\cite{MullerHartmann1994,RaghavanElser1995,PereiraMueller2025}. We streamline the derivation with an embedding trick. It allows for a straightforward derivation of the resummed expression for the current correlation function, which would be challenging with combinatorics, if possible at all. 

We define a tight binding model in a four-coordinated Bethe lattice, $H_B = -t\sum_{rr'}|r\rangle_B {}_B\langle r'|$, where $|r\rangle_B$ is the position eigenstate on the Bethe lattice site $r$. A walk $\Gamma$ on the square lattice can be embedded in the Bethe lattice by translating the string of letters  (Fig.~\ref{fig:bethe_lattice}). In particular, a skeleton $\gamma$ can be identified as a site of the Bethe lattice $r_\gamma$, whereas the hairs are mapped to excursions on the way from the origin to that site.  

We claim:
\begin{align}
w(\gamma) =  {}_B\langle r_\gamma|e^{-\beta H_B}|0\rangle_B.
\label{eq:w}
\end{align}
To prove this, we expand the R.H.S. as $\sum_{\eta} (\beta t)^{n_{\eta}}/n_{\eta}!$. The summation is over all walks $\eta$ in the Bethe lattice that start at the origin 0 and end at $r_\gamma$. We compare this sum with that in the definition of $w(\gamma)$ (Eq.~\eqref{eq:Z_resum}): Each and every $\eta$ is constructed by dressing the unique, shortest path from 0 to $r_\gamma$ in the Bethe lattice with excursions, which is in one-to-one correspondence with the hairs grown out of the skeleton $\gamma$ in the square lattice. Furthermore, the contribution from each walk $\eta$ in the Bethe lattice is identical to the contribution from each dressed path $\Gamma$ in the square lattice. Q.E.D.

\textit{Current correlator ---} We seek the resummed expression for the current correlation function by coupling $H$ to a time-dependent vector potential $A_a(\tau)$:
\begin{align}
H_A(\tau) = -t\sum_{ij \sigma}e^{i\phi_{ij}(\tau)} \mathcal{P} c^\dagger_{i\sigma} c^{\phantom\dagger}_{j\sigma} \mathcal{P}- h \sum_i S^z_i /\hbar.
\end{align} 
The Peierls phase $\phi_{i,i\pm a}(\tau) = \mp e d A_a(\tau)/\hbar$. $a=x,y$. $d$ is the lattice spacing. $e = -|e|$ is the electron charge. The partition function becomes $Z_A = \mathrm{Tr}(\mathcal{T}_\tau e^{-\int^\beta_0 H_A(\tau)d\tau})$, where $\mathcal{T}_\tau$ denotes time ordering. Applying the same analysis to $Z_A$ as in Eq.~\eqref{eq:Z}, we obtain:
\begin{align}
\frac{Z_A}{\Omega} = \sum_{\gamma}f(\gamma)w_A(\gamma),
\label{eq:Z_A_resum}
\end{align}
where the kinetic weight in the gauge field is:
\begin{align}
w_A(\gamma) = \sum_{\Gamma\uparrow\gamma} t^{n_\Gamma} \int^\beta_0 d\tau_{n}e^{-i\phi_{i_ni_{n-1}}(\tau_n)} \cdots 
\nonumber\\
\times \int^{\tau_3}_0 d\tau_2 e^{-i\phi_{i_2i_1}(\tau_2)} \times \int^{\tau_2}_0 d\tau_1 e^{-i\phi_{i_1i_{0}}(\tau_1)}.
\end{align}
The integral is the gauge flux picked up by $\Gamma$. $i_0i_1\cdots i_n$ are successive sites visited by the hole. The minus sign in front of $\phi_{ij}$ reflects the $-e$ charge carried by the hole.

By analogy with the $A_a=0$ case,
\begin{align}
w_A(\gamma) =  _B\langle r_\gamma|\mathcal{T}_\tau e^{-\int^\beta_0 H_{B,A}(\tau)d\tau}|0\rangle_B.
\label{eq:w_A}
\end{align}
$H_{B,A}(\tau) = -t\sum_{rr'}e^{-i\phi_{rr'}(\tau)}|r\rangle_B{}_B\langle r'|$ is the Bethe lattice model in a gauge field. The Peierls phases $\phi_{rr'}(\tau)$ are inherited from the square lattice model. $\phi_{r,r+x/\overline{x}}(\tau) = \mp edA_x(\tau)/\hbar$, and $\phi_{r,r+y/\overline{y}}(\tau) = \mp edA_y(\tau)/\hbar$. The proof of Eq.~\eqref{eq:w_A} is the same as Eq.~\eqref{eq:w}, i.e. expanding both sides as path sums and matching each term in the summation. 

Taking the derivative of $Z_A$ yields the current correlation function: $C_{ab}(\tau) = \delta^2 Z_A/[\delta A_a(\tau) \delta A_b(0)]/Z\big|_{A=0}$ for $0<\tau<\beta$. Using Eq.~\eqref{eq:Z_A_resum} and Eq.~\eqref{eq:w_A}, we find:
\begin{align}
C_{ab}(\tau) = \frac{\sum_{\gamma}f(\gamma) w'_{ab}(\gamma,\tau)}{\sum_\gamma f(\gamma) w(\gamma)}.
\label{eq:current_corr}
\end{align}
The observable for the current correlation function is:
\begin{align}
w'_{ab}(\gamma,\tau) = _B\langle r_\gamma| e^{-(\beta-\tau)H_B} I_a e^{-\tau H_B} I_b |0\rangle_B,
\label{eq:current_resum}
\end{align}
where $I_a = -\partial H_{B,A}/\partial A_a\big|_{A=0}$ is the current operator in the Bethe lattice. Eq.~\eqref{eq:current_corr} and Eq.~\eqref{eq:current_resum} represent the central technical advance of this work.

What remains is a practical method for calculating $w(\gamma)$ and $w'_{ab}(\gamma,\tau)$. We sketch the ideas and refer the readers to the Supplemental Material (SM)~\cite{SupplementalMaterial}\nocite{AryalKettemann2020,Mahan2001,Wolff2004} for an in-depth exposition.

The kinetic weight $w(\gamma)$ represents an imaginary time evolution from the origin to a site $r_\gamma$. Since $H_B$ features massive permutation symmetries about the origin, the evolution is closed in the fully symmetric sector. In this sector, $H_B$ is reduced to an effective, one-dimensional tight-binding Hamiltonian, which is diagonalized analytically. Owing to this symmetry, $w(\gamma)$ depends only on the length of $\gamma$, which is tabulated when required.

The evolution represented by $w'_{ab}(\gamma,\tau)$ is interrupted at time 0 and $\tau$ by the current operators $I_{a,b}$, which mix in less symmetric sectors. As a result, $w'_{ab}$ depends not only on the length of $\gamma$ but also on its shape. Nevertheless, its structure is simple enough for tabulation.

\begin{figure}
\includegraphics[width = \columnwidth]{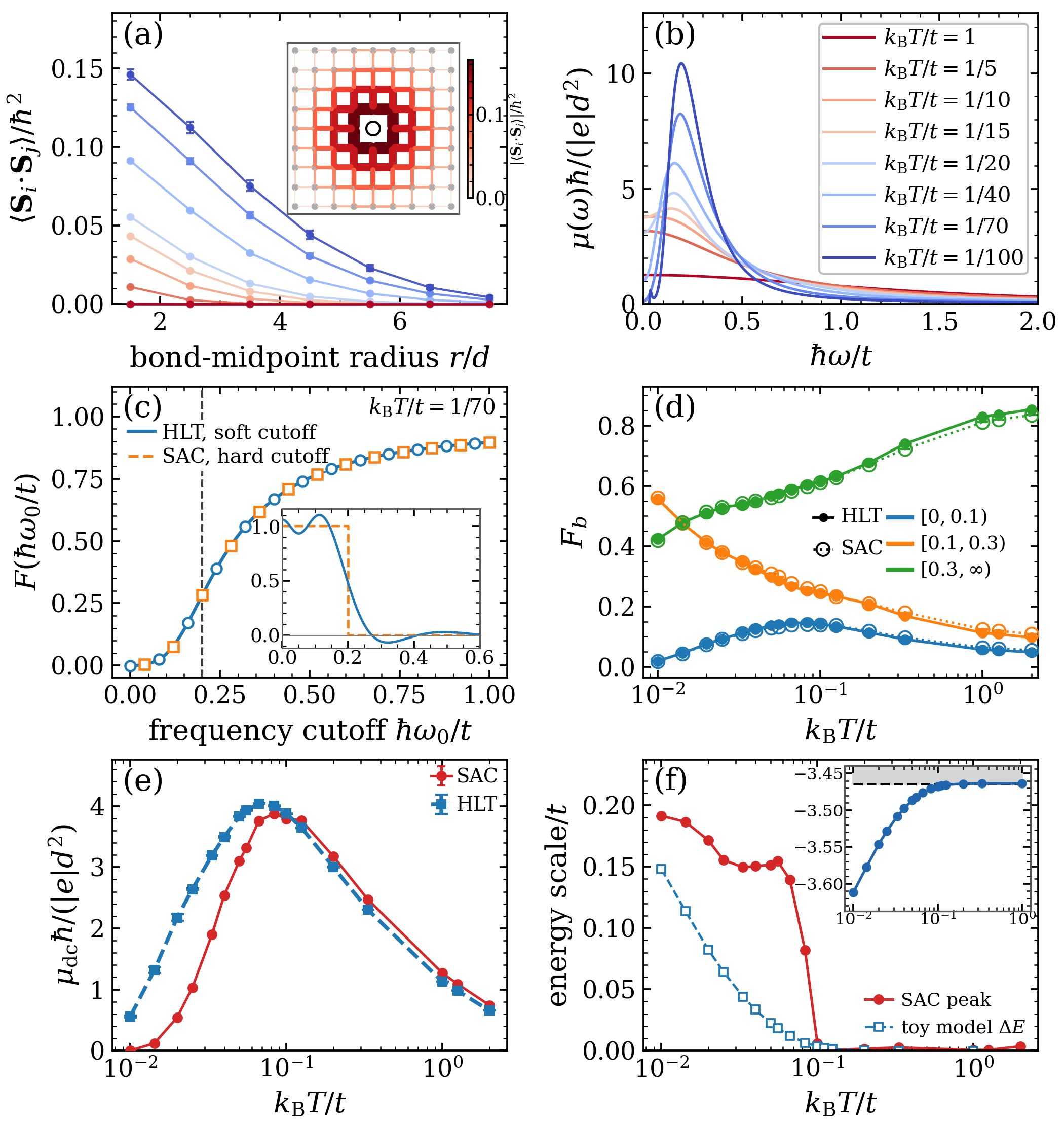}
\caption{Thermal evolution of the dynamical mobility $\mu(\omega)$. (a) Spin correlation on nearest neighbor bonds near the hole cut along the horizontal axis. Inset: spin correlations at $k_BT/t = 1/70$. (b) Temperature driven evolution of $\mu(\omega)$. (c) Cumulative spectral weight at $k_BT/t = 1/70$ obtained by integrating SAC spectrum with a hard cutoff (orange) and by HLT method with a soft cutoff (blue). Inset shows the cutoff functions. (d) Transfer of spectral weights. Filled and open circles are for the HLT and SAC results, respectively. (e) DC mobility estimated from SAC (red) and HLT (blue). (f) Peak position of $\mu(\omega)$ (red) compared with the toy model (blue). Inset shows the binding energies of the hole bound states from the toy model. }
\label{fig:x=0_cut}
\end{figure}

\textit{Numerics ---}  We estimate the imaginary time current correlation function $C_{xx}(\tau)$ (Eq.~\eqref{eq:current_corr}) by Markov chain Monte Carlo. We update $\gamma$ by a set of canonical local moves. See SM~\cite{SupplementalMaterial} for further discussion. We benchmark our algorithm against exact diagonalization on a $4\times4$ system with periodic boundary conditions. Each production run of $C_{ab}(\tau)$ in the thermodynamic limit uses $O(10^3)$ independent Markov chains, with $O(10^8)$ effectively independent samples per chain, resulting in a relative error on the order of $10^{-5}$. The resummation method is increasingly more efficient than the direct method at lower temperature, reaching a $\sim 25$ times speed up at $\beta t = 70$~\cite{SupplementalMaterial}.

We construct the dynamical mobility $\mu(\omega)$ from the current correlation function by SAC for its ability to resolve sharper spectral structures~\cite{Sandvik1998,Shao2023,Sandvik2016}.
The maximum entropy method~\cite{SilverSiviaGubernatis1990,KaufmannHeld2023} reproduces all qualitative features, attesting to the robustness of the continuation~\cite{SupplementalMaterial}. We validate the SAC results by comparing them against the Hansen-Lupo-Tantalo (HLT) method, a continuation free method for constructing spectral observables~\cite{HLT2019}. Starting from,
\begin{align}
C_{xx}(\tau_i) = \int^\infty_0 K(\tau_i,\omega) \mu(\omega) d\omega,
\end{align}
where $K(\tau,\omega)=|e|\hbar\omega\cosh[(\beta/2\!-\!\tau)\hbar\omega]/[\pi\sinh(\beta\hbar\omega/2)]$, HLT constructs a smeared spectral observable by a linear combination of the imaginary time data:
\begin{align}
\sum_i \lambda_i C_{xx}(\tau_i) = \int^\infty_0 f_{HLT}(\omega) \mu(\omega) d\omega.
\end{align}
The filter $f_{HLT}(\omega) = \sum_i \lambda_i K(\tau_i,\omega)$. $\{\lambda_i\}$ are chosen by linear regression such that $f_{HLT}(\omega)$ approximates a target filter $f(\omega)$. We set the target filter $f(\omega) = \Theta(\omega_0-\omega)$, corresponding to (smeared) cumulative spectral weight.

\textit{Nagaoka polaron formation --- }  Fig.~\ref{fig:x=0_cut}(a) shows the smooth buildup of the spin correlations near the hole as the temperature decreases. In sharp contrast to this monotonic behavior, the dynamical mobility $\mu(\omega)$ exhibits a qualitative change: The incoherent Drude-like peak narrows as the temperature $k_BT/t$ decreases from 1 to 0.1, and then a finite frequency peak emerges at lower temperature. Meanwhile, the DC mobility $\mu(0)$ vanishes rapidly, meaning the hole is increasingly localized. $\mu(\omega)$ reveals a fundamental change in the dynamical properties of the hole --- the missing sign for polaron formation.

We validate the result by comparing the cumulative spectral weights obtained from SAC and HLT (Fig.~\ref{fig:x=0_cut}(c)). We obtain the cumulative spectral weight by (a) integrating the SAC spectrum using a hard frequency cutoff, $\int^{\omega_0}_0 \mu_{SAC}(\omega) d\omega$, and (b) estimating it from imaginary time data using a soft cutoff, $\int^\infty_0 f_{HLT}(\omega) \mu(\omega) d\omega$. At the representative temperature $k_BT/t = 1/70$, they agree almost quantitatively. Both increase rapidly at $\hbar\omega_0 \approx 0.2 t$, certifying the finite frequency peak at that scale.

The existence of the finite frequency peak is further corroborated by analyzing the spectral weight transfer (Fig.~\ref{fig:x=0_cut}(d)). We divide the frequency axis $\hbar\omega/t$ into the low ($[0,0.1]$), middle ($[0.1,0.3]$), and high frequency windows ($[0.3,\infty)$). Both SAC and HLT show that the peak arises due to the transfer of spectral weights from both low and high frequencies, suggesting that the peak reflects an internal resonance mode.

We flesh out the internal mode interpretation by using a toy model (Fig.~\ref{fig:phase}(c)). In a fully polarized spin background, the hole propagates as a free particle. When the spins are uncorrelated, it prefers self-retracing walks, reducing the bandwidth by a factor of $\sqrt{3}/2$~\cite{BrinkmanRice1970}. We postulate an effective hopping on bond $ij$, $t_{ij}/t = \sqrt{3}/2+(4-2\sqrt{3})\langle \bm{S}_i\cdot \bm{S}_j\rangle /\hbar^2$, where the coefficients are chosen to match the two limits. The reduction of $t_{ij}$ away from the spin cloud creates a confining potential, which, in turn, binds the hole. The optical transition from the $s$-wave ground state to the first $p$-wave excited state qualitatively tracks the evolution of the $\mu(\omega)$ peak (Fig.~\ref{fig:x=0_cut}(f)), suggesting that the peak reflects the motion of the hole in the confining potential. The toy model shows a smoother thermal evolution due to its mean-field-like character. An earlier droplet picture~\cite{RaghavanElser1995,WhiteAffleck2001} assumes the hole is confined by an infinite potential well, which would predict the \emph{red shift} of the peak as the droplet expands with decreasing temperature, in conflict with the numerically observed \emph{blue shift}.

The DC mobility $\mu(0)$ provides a diagnostic for the Nagaoka polaron. We estimate $\mu(0)$ by using the SAC value $\mu_{SAC}(0)$ and by $\mu(0)\approx \int^{\infty}_0f_{HLT}(\omega) \mu(\omega) d\omega/\omega_0$ where $f_{HLT}$ is a soft cutoff at $\hbar\omega_0 = 0.1t$ (Fig.~\ref{fig:x=0_cut}(e)). As $T$ decreases, both show metal-like increases above $k_BT \approx 0.1t$ and insulator-like decreases below. We take the turning point as the marker for polaron formation, yielding the phase diagram  (Fig.~\ref{fig:phase}(a)). Using the appearance of the finite frequency peak as the marker demarcates a similar crossover.

\begin{figure}
\includegraphics[width=\columnwidth]{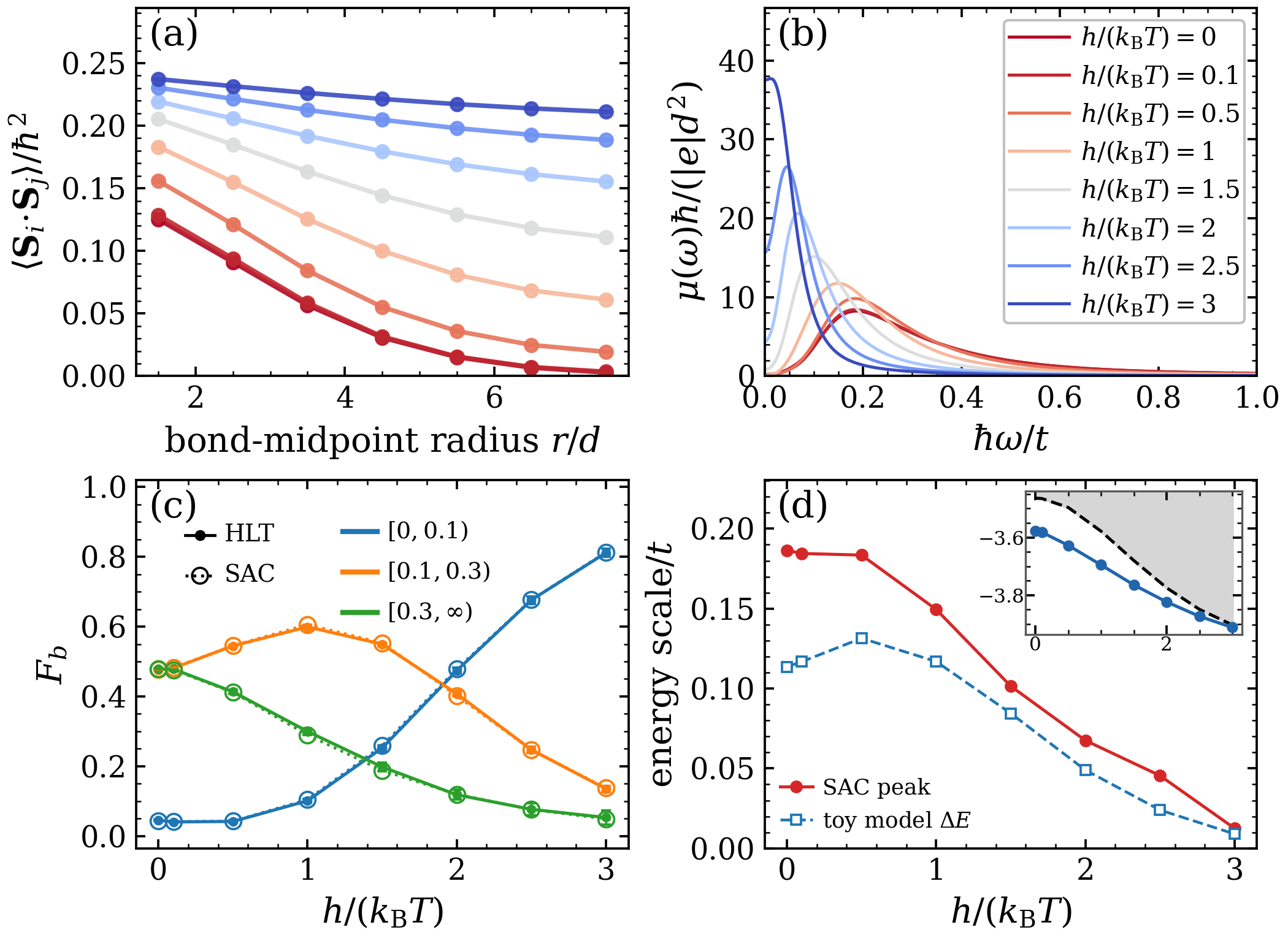}
\caption{Field-driven polaron melting. (a) Spin correlations on nearest neighbor bonds near the hole along the horizontal cut for various Zeeman fields at $k_BT/t = 1/70$. (b) Field-driven evolution of dynamical mobility $\mu(\omega)$ at the same temperature. (c) Transfer of spectral weights constructed from SAC (open circles) and HLT (filled circles). (d) Peak position as a function of field compared against the toy model. Inset: binding energies of bound states from the toy model.}
\label{fig:beta=70_cut}
\end{figure}

\textit{Field-driven melting ---} The previous analysis establishes that the Nagaoka polaron forms by kinetic self-trapping. A Zeeman field suppresses this kinetic mechanism, and, consequently, melts the polaron. At the representative low temperature $k_BT/t = 1/70$, the spin correlations show no qualitative change with the field (Fig.~\ref{fig:beta=70_cut}(a)). However, the finite frequency peak in $\mu(\omega)$ transforms to a Drude-like peak at zero frequency, indicating the Nagaoka polaron melts (Fig.~\ref{fig:beta=70_cut}(b)). Once the polaron melts, the charge transport is dominated by the spin scattering of the hole, which is suppressed by the field. Mirroring this process, the Drude-like peak sharpens as the field increases. SAC and HLT reveal the finite frequency peak disappears by transferring its weight to lower frequency (Fig.~\ref{fig:beta=70_cut}(c)). The field-driven melting picture is further supported by the toy model, namely the field reduces the self-confining potential, resulting in the gradual dissociation of the bound state (Fig.~\ref{fig:beta=70_cut}(d)). 

\textit{Discussion and outlook ---} Our work establishes the dynamical mobility as an incisive probe for the Nagaoka polaron formation. Its rich temperature and field-driven evolution unveils a previously unexplored, dynamical facet of the kinetic magnetism inaccessible to structural diagnostics, thereby opening up a host of new questions such as a quantitative theory for polaron formation, dynamical fingerprints of kinetic frustration~\cite{HaerterShastry2005,Sposetti2014,ZhangBatista2018,Sharma2026}, and incarnations of similar physics in spin liquids~\cite{Hart2020,Hart2021}.

Our approach generalizes to other lattice geometries within the $t$-model. A Monte Carlo sign problem arises for a hole in non-bipartite lattices but remains mild at moderate temperatures~\cite{PereiraMueller2025}, which can be further mitigated by a neural autoregressive control variate~\cite{QiaoWang2026}. We expect the dynamical mobility exhibits a crossover from Drude-like response to finite frequency resonance similar to the unfrustrated case though the temperature scale for the polaron formation and the energy scale of the resonance could be parametrically different. Exploring transport signatures of kinetic frustration will be a promising direction. Introducing a weak orbital magnetic field allows us to explore the magnetotransport of Nagaoka polarons~\cite{BrinkmanRiceHall1971,Lange1999}. Finally, adapting our approach to the thermal transport~\cite{OguriMaekawa1990} and the finite doping regime presents important open challenges.

\begin{acknowledgments}
We thank Xinyang Dong for discussions and Darren Pereira for comments on the manuscript. This work is supported by the National Key R\&D Program of China (Grants No. 2022YFA1403800, 2024YFA1408700), the National Science Foundation of China (Grants No. 12547112, 12188101), and the CAS Project for Young Scientists in Basic Research (Grant No. YSBR-059).
\end{acknowledgments}

\clearpage
\onecolumngrid
% Supplemental counters and anchors are independent of the main text.
% REVTeX assumes a single bibliography. Give its final-item label a separate
% name here, and do not print the main title's author footnotes a second time.
\makeatletter
\def\NAT@bibsetnum#1{%
  \setlength{\topsep}{0pt}%
  \NATx@bibsetnum{\ref{sm:LastBibItem}}%
}
\renewenvironment{thebibliography}[1]{\NAT@thebibliography{#1}}{%
  \edef\@currentlabel{\arabic{NAT@ctr}}%
  \label{sm:LastBibItem}%
  \endNAT@thebibliography
  \aftergroup\auto@bib@empty
}
\makeatother
\setcounter{section}{0}
\setcounter{subsection}{0}
\setcounter{equation}{0}
\setcounter{figure}{0}
\setcounter{table}{0}
\setcounter{secnumdepth}{3}
\renewcommand{\thesection}{\Roman{section}}
\renewcommand{\thesubsection}{\Alph{subsection}}
\renewcommand{\theequation}{S\arabic{equation}}
\renewcommand{\thefigure}{S\arabic{figure}}
\renewcommand{\thetable}{S\arabic{table}}
\renewcommand{\theHsection}{supp.\arabic{section}}
\renewcommand{\theHequation}{supp.\arabic{equation}}
\renewcommand{\theHfigure}{supp.\arabic{figure}}
\renewcommand{\theHtable}{supp.\arabic{table}}
\phantomsection
\label{sec:supplement}
\pdfbookmark[0]{Supplemental Material}{supplement}
\begin{center}
{\bfseries Supplemental Material for\\
``Transport Signatures of a Kinetically Self-Trapped Nagaoka Polaron''\par}
\medskip
Yusong Xiao$^{1,2}$ and Yuan Wan$^{1,3}$\par
\smallskip
{\small\itshape
$^1$Institute of Physics, Chinese Academy of Sciences, Beijing 100190, China\\
$^2$University of Chinese Academy of Sciences, Beijing 100049, China\\
$^3$Songshan Lake Materials Laboratory, Dongguan, Guangdong 523808, China\par}
\smallskip
{\small\href{mailto:yuan.wan@iphy.ac.cn}{yuan.wan@iphy.ac.cn}\par}
\end{center}
\begin{quotation}\small
This document provides additional information on the calculation of the auxiliary Bethe lattice quantities, the quantum Monte Carlo algorithm, and the numerical analytic continuation.
\end{quotation}

\section{Auxiliary Bethe lattice quantities}

In the main text, we have defined the tight-binding model in the coordination-four Bethe lattice:
\begin{align}
H_B=-t\sum_{rr'}|r\rangle_B{}_B\langle r'|,
\label{sm:eq:Hamiltonian}
\end{align}
$|r\rangle_B$ is the position eigenstate on the Bethe lattice site $r$. The summation is over all oriented nearest neighbor links $rr'$. $t>0$ is the hopping amplitude. The resummation-based quantum Monte Carlo requires the following objects:
\begin{subequations}
\begin{align}
w(r)&={}_B\langle r|e^{-\beta H_B}|0\rangle_B;
\\
w'_{xx}(r,\tau) &={}_B\langle r|e^{-(\beta-\tau)H_B}I_xe^{-\tau H_B}I_x|0\rangle_B.
\end{align}
\label{sm:app:eq:aux_bethe}
\end{subequations}
$I_x$ is the current operator in the Bethe lattice. In this section, we evaluate these auxiliary Bethe lattice objects. Our strategy is block diagonalizing $H_B$ in a symmetry-adapted basis~\cite{sm:AryalKettemann2020}, and then calculating $w(r)$ and $w'_{xx}(r,\tau)$ in that basis. 

\subsection{Symmetry-adapted basis}

We provide a descriptive construction of the symmetry-adapted basis in the Bethe lattice~\cite{sm:AryalKettemann2020}. We do not touch the representation theory aspects of this construction as it requires an extensive discussion of group theory materials. Instead, we illustrate the usefulness of this basis by examining the action of $H_B$ in this basis.

Starting from the origin $0$, each site $r$ in the Bethe lattice is uniquely associated with the shortest walk from the origin to $r$, which, in turn, is parametrized as a string of letters:  $r=s_1\cdots s_l$. The letter $s_j$, taken from the set $\{x,\bar{x},y,\bar{y}\}$, denotes the direction of the $j$-th step. Furthermore, the walk admits no immediate reversal, $s_{j+1}\neq\bar s_j$. We call $l$ the level of the site $r$. Specifically, the number of sites at level $l$ is $N_0 = 1$ and 
$N_l = 4\times 3^{l-1}$ for $l\ge 1$. 

We may recombine all position eigenstates at that level, $|s_1\cdots s_l\rangle_B$, to form appropriate symmetry-adapted basis states. We assume $l\ge 2$ for now. First, we consider the 3 level-$l$ states $|s_1\cdots s_{l-1}a\rangle_B$, with $a\neq\overline{s}_{l-1}$, sharing the same parent $s_1\cdots s_{l-1}$. They transform among themselves under the $S_3$ permutations with respect to the parent. We form the one-dimensional irreducible representation of the $S_3$ group:
\begin{align}
|s_1\cdots s_{l-1}\psi_A\rangle_B = \frac{1}{\sqrt{3}}\sum_{a\neq \overline{s}_{l-1}} |s_1\cdots s_{l-1}a\rangle_B,
\end{align}
and the two-dimensional irreducible representation:
\begin{align}
|s_1\cdots s_{l-1}\psi^{(i)}_E\rangle_B = \frac{1}{\sqrt{3}}\sum_{a\neq \overline{s}_{l-1}} \psi^{(i)}_E(a)|s_1\cdots s_{l-1}a\rangle_B.
\end{align}
Here, $\psi^{(i)}_E$ are the two basis vectors that carry the two-dimensional representation of $S_3$.

Second, we set the $|s_1\cdots s_{l-1}\psi^{(i)}_E\rangle$ states aside, and focus on the symmetric combination $|s_1\cdots s_{l-1}\psi_A\rangle$. Consider three such states that share the same parent, $|s_1\cdots s_{l-2}\alpha \psi_A\rangle$, with $\alpha\neq \overline{s}_{l-2}$. They transform among themselves under the $S_3$ permutations with respect to the parent $s_1\cdots s_{l-2}$. We form a one-dimensional irreducible representation of the $S_3$ group: 
\begin{align}
|s_1\cdots s_{l-2}\psi_A\psi_A\rangle_B = \frac{1}{\sqrt{3}}\sum_{a\neq \overline{s}_{l-2}} |s_1\cdots s_{l-2}a \psi_A\rangle_B,
\end{align}
and two-dimensional irreducible representation:
\begin{align}
|s_1\cdots s_{l-2}\psi^{(i)}_E\psi_A\rangle_B = \frac{1}{\sqrt{3}}\sum_{a\neq \overline{s}_{l-2}} \psi^{(i)}_E(a)|s_1\cdots s_{l-2}a \psi_A\rangle_B.
\end{align}

Repeating the above process, we obtain states of the form:
\begin{align}
|s_1\cdots s_p \psi^{(i)}_E \underbrace{\psi_A\cdots \psi_A}_{l-p-1}\rangle_B, \quad (p\ge 1).
\end{align}
The remaining states are:
\begin{align}
|s_1 \underbrace{\psi_A\cdots \psi_A}_{l-1}\rangle_B.
\end{align}
These states transform among themselves under the $S_4$ permutation about the origin (root). We may form a one-dimensional irreducible representation of the $S_4$ group:
\begin{align}
|\underbrace{\psi_A\cdots \psi_A}_{l}\rangle_B = \frac{1}{\sqrt{4}}\sum_{s_1} |s_1 \underbrace{\psi_A\cdots \psi_A}_{l-1}\rangle_B,
\end{align}
and the three-dimensional irreducible representation:
\begin{align}
|\psi^{(i)}_T \underbrace{\psi_A\cdots \psi_A}_{l-1}\rangle_B = \frac{1}{\sqrt{4}}\sum_{s_1} \psi^{(i)}_T(s_1) |s_1 \underbrace{\psi_A\cdots \psi_A}_{l-1}\rangle_B,
\end{align}
where $\psi^{(i)}_T$ are the basis vectors that carry the three-dimensional representation of $S_4$.

\begin{figure}
\includegraphics[width=\columnwidth]{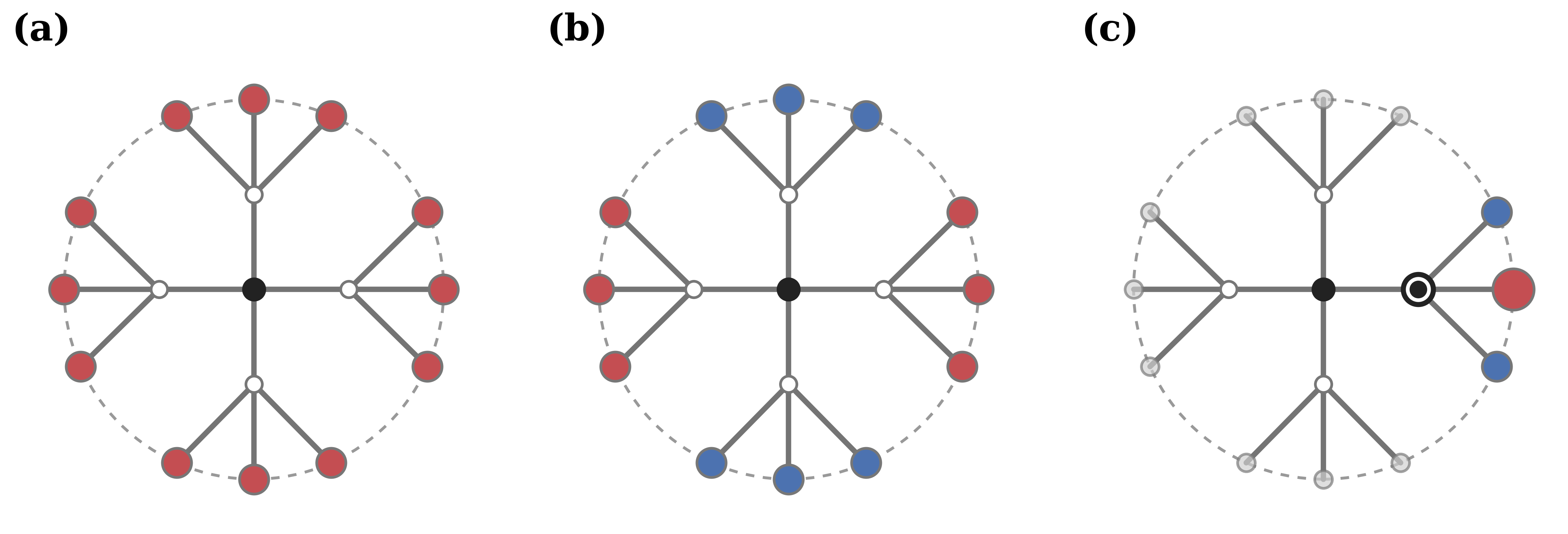}
\caption{Representative symmetry-adapted basis states. Red, blue, and gray circles denote positive, negative, and zero amplitudes, respectively. The radius of the circle indicates the magnitude of the amplitude. (a) Radial state $|2\rangle_B$, with uniform amplitude on the second shell. (b) Root-angular state $|\psi_T,1\rangle_B$. (c) Node-angular state $|x,\psi_E,0\rangle_B$.}
\label{sm:fig:symmetry-adapted-basis}
\end{figure}

The above construction assumes the level $l\ge 2$. It can be extended to $l=1$ and $l=0$. When $l=0$, there is only one state, $|0\rangle_B$. For $l=1$, there are one symmetric combination $|\psi_A\rangle_B$ and three asymmetric combinations $|\psi^{(i)}_T\rangle_B$.

To summarize, we obtain the following families of basis states:
\begin{itemize}
\item Radial states:
\begin{align}
|l\rangle_B = |\underbrace{\psi_A\cdots \psi_A}_{l}\rangle_B = \frac{1}{\sqrt{N_l}}\sum_{s_1\cdots s_l} |s_1\cdots s_l\rangle_B.
\end{align}
These states are fully symmetric, and are well defined for any $l\ge 0$. In particular, $|0\rangle_B$ belongs to this family.

\item Root angular states:
\begin{align}
|\psi^{(i)}_T, l-1\rangle_B = |\psi^{(i)}_T \underbrace{\psi_A\cdots \psi_A}_{l-1}\rangle_B =  \frac{1}{\sqrt{4\times3^{l-1}}}\sum_{s_1\cdots s_l} \psi^{(i)}_T(s_1) |s_1 \cdots s_l\rangle_B.
\end{align}
They are defined for $l\ge 1$. The root angular states carry finite ``angular momentum" about the origin (root). Their amplitudes on the origin vanish by symmetry.

\item Node angular state:
\begin{align}
|s_1\cdots s_p,\psi^{(i)}_E,l-p-1\rangle_B = |s_1\cdots s_p \psi^{(i)}_E \underbrace{\psi_A\cdots \psi_A}_{l-p-1}\rangle_B = \frac{1}{\sqrt{3^{l-p}}}\sum_{s_{p+1} \cdots s_l} \psi^{(i)}_E(s_{p+1}) |s_1\cdots s_p s_{p+1} \cdots s_l\rangle_B.
\end{align}
Such states are defined for $1\le p \le l-1$. For a given ``anchor" $s_1\cdots s_p$, the node angular states are localized to the children branches of the anchor site, and carry finite angular momentum about the anchor. 
\end{itemize}
It is easy to see that the basis states are orthogonal by symmetry. Furthermore, they span the entire one-particle Hilbert space. To see this, we count the number of symmetry-adapted basis states of level $l\ge 2$:
\begin{align}
1 + 3 + \sum^{l-1}_{p=1}(2N_p) = 4\times 3^{l-1} = N_l.
\end{align}
Fig~\ref{sm:fig:symmetry-adapted-basis} shows representative symmetry-adapted basis states with level $l=2$. 

\subsection{Block diagonalization}

$H_B$ is block diagonal in the symmetry-adapted basis. To this end, it is sufficient to examine the action of $H_B$ on these basis states. For the radial states,
\begin{subequations}
\begin{align}
H_B|0\rangle_B &= -2t|1\rangle_B;
\\
H_B|1\rangle_B &=-2t|0\rangle_B - \sqrt3\,t|2\rangle_B;
\\
H_B|n\rangle_B &=-\sqrt3\,t (|n-1\rangle_B +|n+1\rangle_B ), \quad (n\geq2).
\end{align}
\end{subequations}
The action of $H_B$ is closed in the subspace spanned by the radial states $|n\ge0\rangle_B$. In this subspace, $H_B$ is reduced to an effective tight-binding Hamiltonian in a semi-infinite chain, where $n$ labels the sites of the chain. 

For the root-angular states, 
\begin{subequations}
\begin{align}
H_B|\psi^{(i)}_T,0\rangle_B &=-\sqrt3\,t|\psi^{(i)}_T,1\rangle_B ;
\\
H_B|\psi^{(i)}_T,n\rangle_B &=-\sqrt3\,t (|\psi^{(i)}_T, n-1\rangle_B +|\psi^{(i)}_T, n+1\rangle_B ), \quad (n \geq1).
\end{align}
\end{subequations}
The action of $H_B$ is closed in the subspace spanned by $|\psi^{(i)}_{T},n\ge 0\rangle_B$. Its matrix elements are independent of the vector $\psi^{(i)}_T$. $H_B$ is reduced to an effective tight-binding Hamiltonian in a semi-infinite chain in this subspace similar to the case with radial states.

Finally, the action of $H_B$ on the node-angular states is almost identical to that of the root-angular states:
\begin{subequations}
\begin{align}
H_B|v, \psi^{(i)}_E,0\rangle_B &=-\sqrt3\,t|v,\psi^{(i)}_E,1\rangle_B;
\\
H_B|v, \psi^{(i)}_E,n\rangle_B &=-\sqrt3\,t (|v,\psi^{(i)}_E, n-1\rangle_B + |v,\psi^{(i)}_E, n+1\rangle_B ), \quad (n \geq1).
\end{align}
\end{subequations}
$v$ denotes an anchor site. We note that the matrix elements of $H_B$ are independent of the vector $\psi^{(i)}_E$ and the anchor position $v$.

It is convenient for our purpose to define the radial-sector propagator:
\begin{align}
g_{mn}(u)\equiv {}_B\langle m|e^{-uH_B}|n\rangle_B,
\end{align}
and the root-angular/node-angular sector propagator 
\begin{align}
g'_{mn}(u) \equiv {}_B\langle \psi^{(i)}_T,m|e^{-uH_B}|\psi^{(i)}_T, n\rangle_B = {}_B\langle v, \psi^{(i)}_E,m|e^{-uH_B}|v,\psi^{(i)}_E,n\rangle_B.
\end{align}
Note $g'$ is independent of $\psi^{i}_T$, $\psi^{i}_E$, or $v$ owing to the fact that the matrix elements of $H_B$ in the root/node-angular subspaces are independent of these objects. 

We calculate $g_{mn}$ and $g'_{mn}$ by first diagonalizing the effective one-dimensional tight binding Hamiltonian in the standing-wave basis~\cite{sm:Mahan2001}, and then expressing the resulting wave vector integral as special functions: \begin{align}
g_{mn}(u) &= \left\{\begin{array}{ll}
D_0(tu) & (m=n=0) \\
\dfrac{2}{\sqrt3} D_{m+n}(tu) & (m=0,n\neq0\enskip\textrm{or}\enskip n=0,m\ne0) \\
D_{m+n}(tu)+\dfrac13D_{|m-n|}(tu)
+\dfrac23\displaystyle\sum_{d=1}^{\min(m,n)}D_{m+n-2d}(tu) & (m\neq 0,n\neq 0)
\end{array}\right. .
\\
g'_{mn}(u) &= I_{|m-n|}(2\sqrt3\,tu)-I_{m+n+2}(2\sqrt3 tu).
\end{align}
Here, $I_k$ is the modified Bessel function of the first kind. $D_{d\ge0}(x)$ is an auxiliary function:
\begin{align}
D_d(x)=I_d(2\sqrt3\,x)-2\sum_{j=1}^{\infty}3^{-j}I_{d+2j}(2\sqrt3\,x).
\end{align} 

\subsection{Resummed observables}

We are now ready to calculate the auxiliary Bethe lattice quantities defined in Eq.~\eqref{sm:app:eq:aux_bethe}. We first consider the kinetic weight $w(r)$. As $|0\rangle_B$ belongs to the radial sector, the ensuing evolution induced by $\exp(-\beta H_B)$ is restricted to this sector. We thus find:
\begin{align}
w(r)=\frac{1}{\sqrt{N_l}} {}_B\langle l|e^{-\beta H_B}|0\rangle_B =\frac{g_{l0}(\beta)}{\sqrt{N_l}},
\end{align}
where $l$ is the level of the Bethe lattice site $r$. We see that $w$ depends on $r$ only through its level, i.e. $w(r) = w(l)$. In practice, we precompute a list of $w(l)$ and tabulate them for later Monte Carlo use.

We turn to the resummed observable $w'_{xx}(r)$. The matrix elements of  $I_x$, the current operator in the Bethe lattice, are conveniently expressed through its action on a Bethe lattice position eigenstate $|r\rangle_B$. Writing $r = s_1\cdots s_l$:
\begin{align}
I_x|s_1\cdots s_l \rangle_B = i\frac{|e|dt}{\hbar} \left\{ \begin{array}{cc}
|s_1\cdots s_l x\rangle_B - |s_1\cdots s_l \overline{x}\rangle_B & (s_l\neq x,\overline{x}) \\
|s_1\cdots x x\rangle_B - |s_1\cdots s_{l-1} \rangle_B & (s_l= x) \\
|s_1\cdots s_{l-1}\rangle_B - |s_1\cdots s_{l-1}\overline{xx}\rangle_B & (s_l=\overline{x}) \\ 
\end{array} \right. .
\end{align}
$|e|$ is the hole charge. $d$ is the lattice spacing. $t$ is the hopping amplitude.  In particular,
\begin{align}
I_x|0\rangle_B = i\frac{|e|dt}{\hbar} (|x\rangle_B - |\overline{x}\rangle_B).
\end{align}
It is clear that $I_x$ mixes the symmetry-adapted basis states in different symmetry sectors. Nevertheless, by inserting the resolution of the identity, we find the resummed observable admits the following decomposition:
\begin{subequations}
\begin{align}
w'_{xx}(r,\tau) = w'_\mathrm{rad}(r,\tau) + w'_\mathrm{root-ang}(r,\tau) + w'_\mathrm{node-ang}(r,\tau).
\end{align}
Writing $r = s_1\cdots s_l$, the radial contribution only depends on $l$:
\begin{align}
w'_\mathrm{rad} (r,\tau) = \frac{(edt)^2}{\hbar^2} \frac{1}{\sqrt{N_l}}\sum_{n=0}^{\infty}g'_{n0}(\tau) [(3^{1/2-n}+(2-\sqrt3)\delta_{n0}) g_{nl}(\beta-\tau)-3^{-n-1/2}g_{n+2,l}(\beta-\tau)] = w'_\mathrm{rad}(l,\tau).
\end{align}
The root-angular contribution depends on $l$ and the first letter $s_1$ (the direction of the first step):
\begin{align}
w'_\mathrm{root-ang}(r,\tau) =  \frac{(edt)^2}{\hbar^2}  \Gamma_T^x(s_1)  3^{-l/2} \sum_{m=0}^{\infty}3^{-m} [g'_{m,l-1}(\beta-\tau) g'_{m+1,0}(\tau)-g'_{m+1,l-1}(\beta-\tau)g'_{m0}(\tau) ] 
\nonumber\\
= \Gamma_T^x(s_1) \widetilde{w}'_\mathrm{root-ang}(l,\tau).
\end{align}
In the second line, we separate the part that depends on the first letter $s_1$ and the part that depends on $l$. The node-angular contribution is more involved:
\begin{align}
w'_\mathrm{node-ang}(r,\tau) = \frac{(edt)^2}{\hbar^2}  \eta_x(s_1)3^{-l/2}\sum_{p=2}^l \sum_{m=0}^{\infty}3^{-m}
\Gamma_E^x(s_{p-1},s_p) g'_{m,l-p}(\beta-\tau) [g'_{m+p,0}(\tau)-3g'_{m+p-2,0}(\tau) ]
\nonumber\\
= \eta_x(s_1) \sum^l_{p=2}\Gamma_E^x(s_{p-1},s_p) \widetilde{w}'_{node-ang}(p,l,\tau).
\end{align}
\label{sm:app:eq:resum_obs}
\end{subequations}
Here, we have defined a set of symmetry-related factors:
\begin{subequations}
\begin{align}
\Gamma_T^x(s) &= \delta_{s,x}+\delta_{s,\bar x}-\frac12,
\\
\eta_x(s) &= \delta_{s,x}-\delta_{s,\bar x},
\\
\Gamma_E^x(s,s') &= \eta_x(s')-\frac13\eta_x(s).
\end{align}
\end{subequations}
Eq.~\eqref{sm:app:eq:resum_obs} provides an efficient way of calculating the resummed observable $w'_{xx}(r)$ for any given $r$. In practice, we precompute and tabulate $w'_{rad}(l,\tau)$, $\widetilde{w}'_{root-ang}(l,\tau)$, and $\widetilde{w}'_{node-ang}(p,l,\tau)$. The resummed observable $w'_{xx}(r,\tau)$ is then calculated on the fly.

\section{Monte Carlo algorithm}

This section describes the local Monte Carlo updates, benchmarks the results against exact diagonalization, and compares the performance of resummed and direct sampling.

\subsection{Algorithm}

\begin{figure}
\includegraphics[width=\columnwidth]{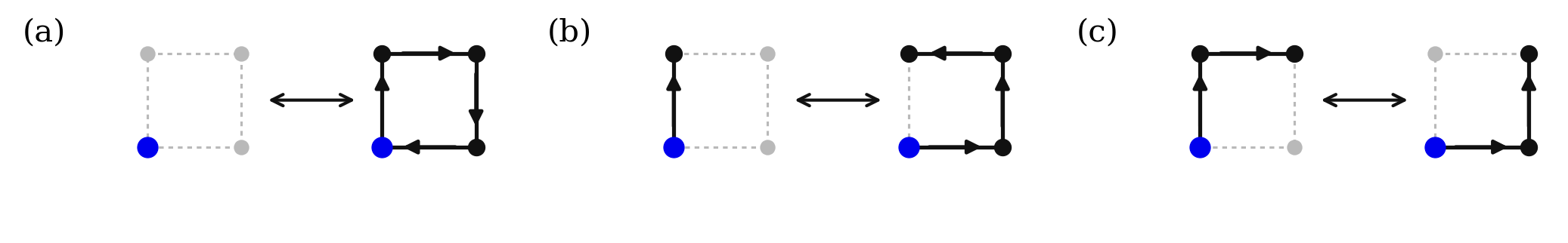}
\caption{Elementary Monte Carlo updates. Black solid elements denote the current path, gray dashed elements indicate the alternative path involved in the proposal, blue markers denote the selected vertex, small arrows indicate the path direction, and double arrows indicate reversible updates. (a) Plaquette insertion and deletion. (b) Detour insertion and deletion. (c) Adjacent-step swap. The three panels are adapted from Figs.~3(a), 3(c), and 3(e) of Ref.~\cite{sm:PereiraMueller2025}.}
\label{sm:fig:mc-updates}
\end{figure}

\begin{table}
\begin{tabular}{ccccc}
\hline
$p^+_\mathrm{plaq}$ & $p^-_\mathrm{plaq}$ & $p^+_\mathrm{detour}$ & $p^-_\mathrm{detour}$ & $p_\mathrm{swap}$ \\
\hline
0.025 & 0.003125 & 0.35 & 0.175 & 0.446875 \\
\hline
\end{tabular}
\caption{Assigned probabilities for local moves.}
\label{sm:tab:update_probability}
\end{table}

We sample the closed skeleton walks $\gamma$ from the distribution:
\begin{align}
P(\gamma) \sim f(\gamma)w(\gamma),
\end{align}
using the Metropolis-Hastings algorithm. $f$ is the spin factor; $w$ is the kinetic weight. We update $\gamma$ with a set of local moves (Fig.~\ref{sm:fig:mc-updates}): inserting/deleting a square shaped plaquette, inserting/deleting a detour, and swapping~\cite{sm:PereiraMueller2025}.  An update begins by selecting one of the five moves according to its assigned probability. For the plaquette insertion, we choose uniformly among the eight plaquettes touching a given square lattice site. Likewise, for the detour insertion, we choose uniformly between the two perpendicular detour directions. A move that is inapplicable or results in a non-skeleton walk is rejected immediately. Detailed balance requires $p_{\mathrm{plaq}}^{+}:p_{\mathrm{plaq}}^{-}=8:1$ and $p_{\mathrm{det}}^{+}:p_{\mathrm{det}}^{-}=2:1$, where $+$ and $-$ denote insertion and deletion. Our choice is given in Tab.~\ref{sm:tab:update_probability}.

Each production run of the current correlation function $C_{xx}(\tau)$ uses 1120 independent Markov chains. Each chain is equilibrated for $4\times 10^5$ Monte Carlo steps (MCS). After equilibration, we estimate the integrated autocorrelation time $\tau_{\mathrm{int}}$~\cite{sm:Wolff2004}, and discard $\tau_{\mathrm{int}}$ MCS between two samples. For each chain, we divide the samples into $N_{\mathrm{block}}=615$ blocks, each block consisting of $2^{16}$ samples. The expectation values are estimated from the block averages, whose fluctuations determine the statistical errors and the covariance matrices used for analytic continuation.

\subsection{Benchmark}

\begin{figure}
\includegraphics[width=\columnwidth]{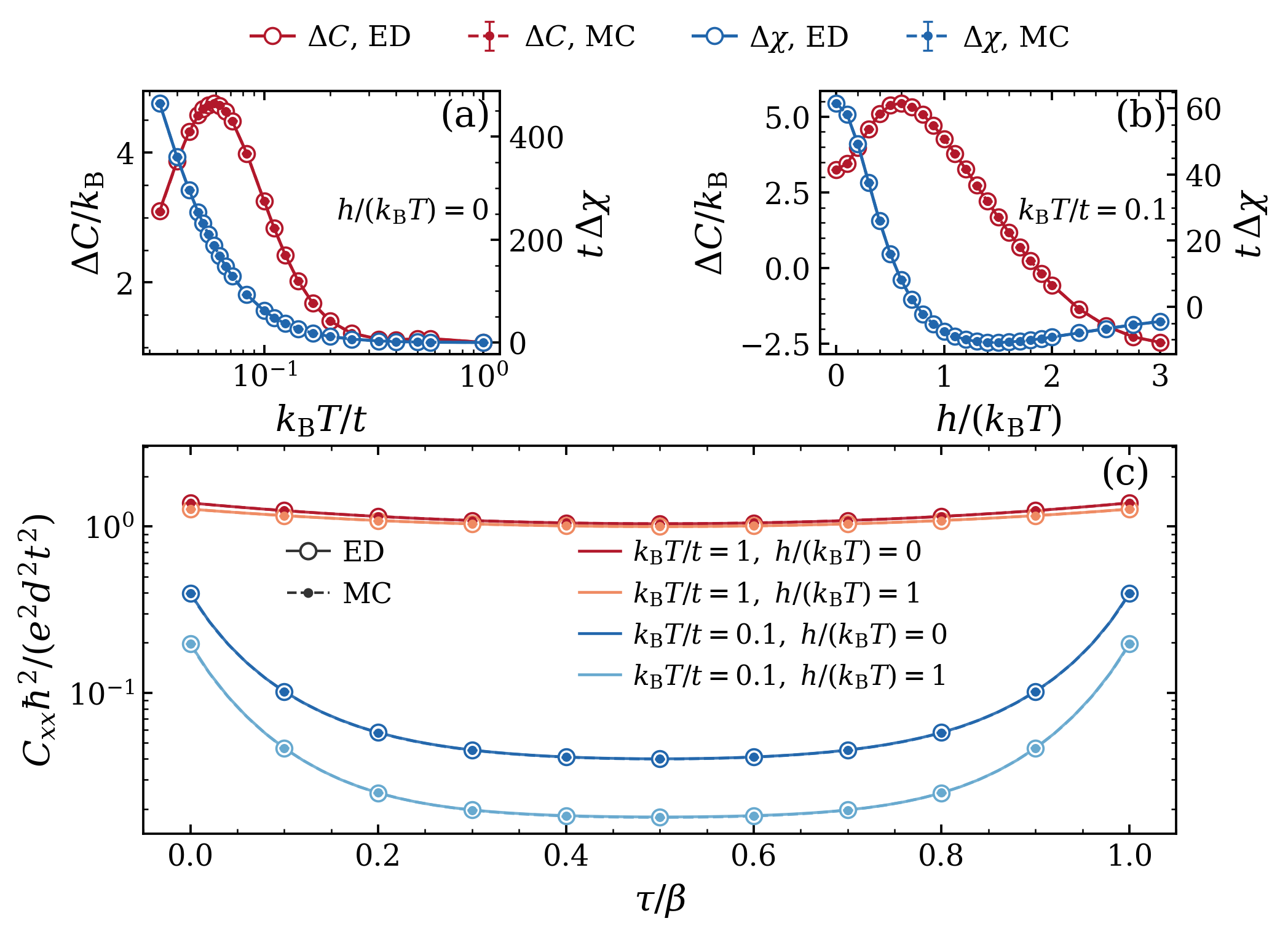}
\caption{Benchmark of the quantum Monte Carlo method against exact diagonalization of a $4\times4$ cluster with periodic boundary conditions. Open circles denote exact diagonalization results. Filled circles denote Monte Carlo results. The error bars are smaller than the size of the symbol. (a) Excess heat capacity $\Delta C$ and excess magnetic susceptibility $\Delta\chi$ as functions of temperature at zero field $h=0$. (b) $\Delta C$ and $\Delta \chi$ as functions of the scaled magnetic field $x=h/(k_BT)$ at $k_BT/t=0.1$. (c) Imaginary time correlation function $C_{xx}(\tau)$ for various choices of temperatures and fields.}
\label{sm:fig:benchmark}
\end{figure}

We benchmark our resummation-based quantum Monte Carlo algorithm against exact diagonalization of a $4\times4$ cluster with periodic boundary conditions. Note our algorithm is formulated directly in the thermodynamic limit. The periodic boundary conditions are implemented by identifying square lattice sites with the equivalence relation: $(x,y) \cong (x+4,y)$ and $(x,y) \cong (x,y+4)$. To allow transitions between winding sectors, we supplement the local moves with a pair of updates that insert or delete four consecutive steps in the same lattice direction. Such a segment forms a closed loop winding once around the periodic cluster. Fig.~\ref{sm:fig:benchmark} shows the comparison of various key observables obtained from the Monte Carlo and from the exact diagonalization. Across all panels, the Monte Carlo results agree with the exact diagonalization within the statistical uncertainties; the differences are indiscernible on the scale of the figure.

\subsection{Performance}

\begin{figure}
\includegraphics[width=\columnwidth]{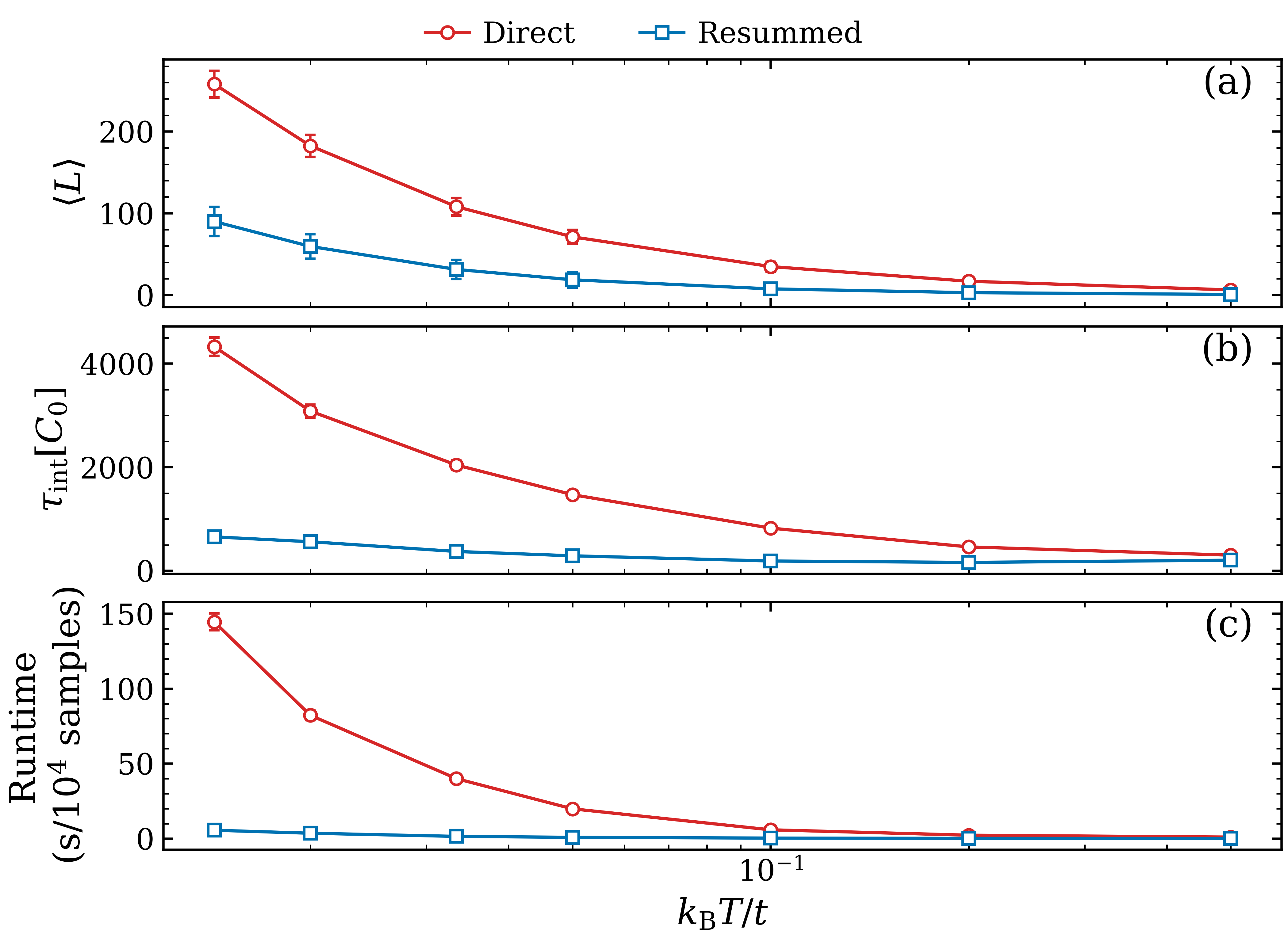}
\caption{Comparison of the performance of the resummation method (blue open squares) and the direct method (red open circles). (a) Mean path length $\langle L\rangle$. (b) Integrated autocorrelation time $\tau_{\mathrm{int}}$ of the current correlation function at Matsubara frequency $i\omega_n = 0$. (c) Wall-clock time per $10^4$ effectively independent samples.}
\label{sm:fig:performance}
\end{figure}

We compare the performance of the resummation method and the direct method in Fig.~\ref{sm:fig:performance}. In the direct method, we sample all closed walks $\Gamma$ in the square lattice from the distribution:
\begin{align}
P(\Gamma) \sim f(\Gamma)\frac{(\beta t)^{n_\Gamma}}{n_\Gamma!},
\end{align} 
and compute various observables accordingly. 

Fig.~\ref{sm:fig:performance}(a) shows that the length of the skeleton $\gamma$ is parametrically shorter than the bare path $\Gamma$. As the computational cost for each Monte Carlo step roughly scales with the length of the walk, the first advantage in the resummation method is the reduced computational cost.

The second advantage for the resummation method is the reduced autocorrelation (Fig.~\ref{sm:fig:performance}(b)). We may understand it as follows. In the direct method, the Markov chain spends most of its time exploring the self-retracing walks, which are strongly correlated with each other. The resummed method instead averages over many walks that share the same skeleton path, allowing the chain to explore distinct  configurations more efficiently.

These two gains act multiplicatively rather than independently. Fig.~\ref{sm:fig:performance}(c) shows the wall-clock time for generating $10^4$ effectively independent samples, separated by $\tau_{int}$ Monte Carlo steps. The resummed method becomes progressively more efficient than the direct method as the temperature decreases. At $\beta t = 70$, the resummation renders $\sim 25$ times speed up. The high performance of the resummation method is essential for obtaining high quality current correlation function data for analytic continuation at low temperature.

\section{Analytic continuation}

\begin{figure}
\includegraphics[width=\columnwidth]{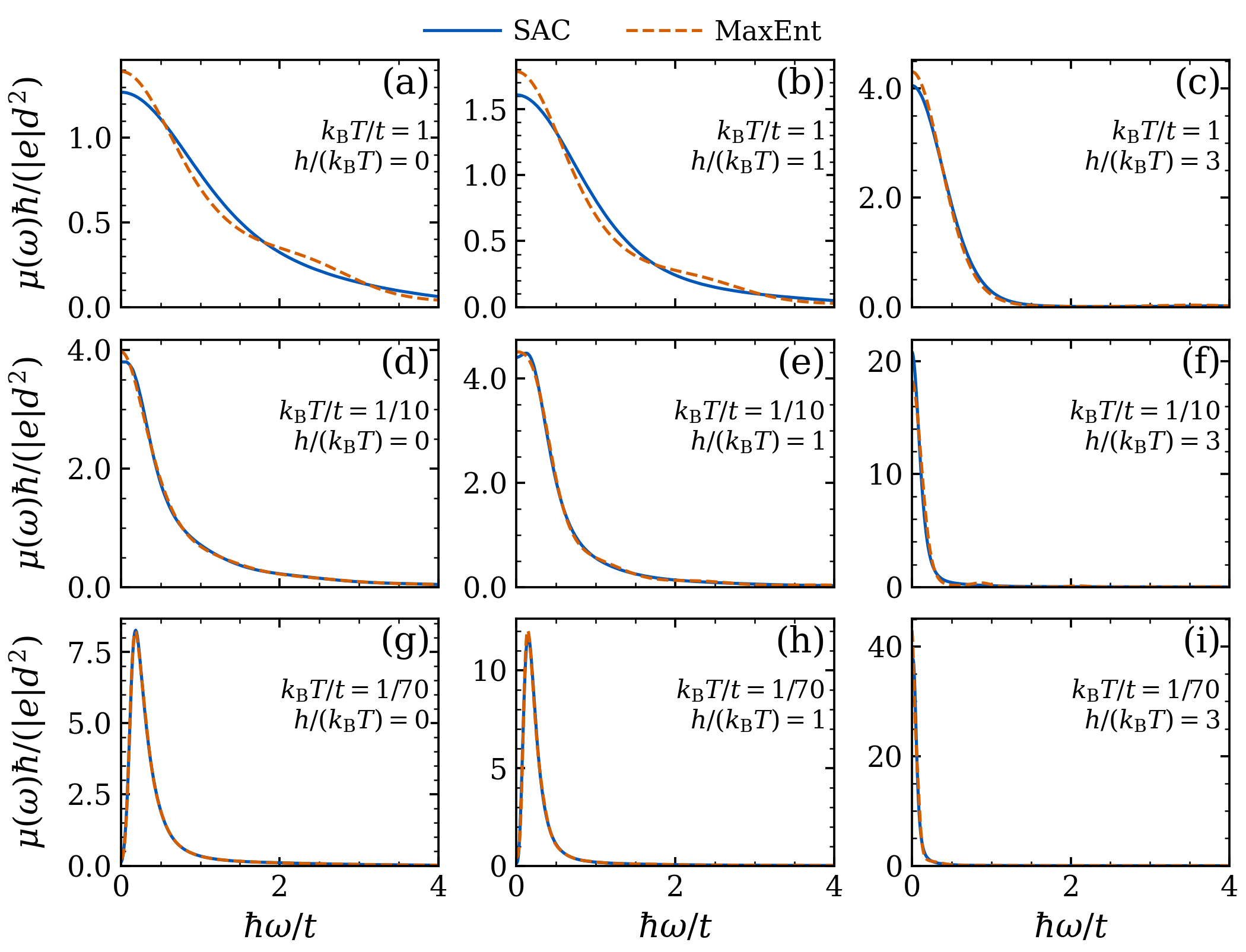}
\caption{Comparison of the dynamical mobility $\mu(\omega)$ constructed by stochastic analytic continuation (SAC, solid blue) and maximum entropy (MaxEnt, dashed orange). The temperature $k_{\mathrm B}T/t=1$, $1/10$, and $1/70$ for the top, middle, and bottom rows. The scaled field $x=h/(k_{\mathrm B}T)=0$, $1$, and $3$ for the left, middle, and right columns.}
\label{sm:fig:continuation-comparison}
\end{figure}

Fig.~\ref{sm:fig:continuation-comparison} compares the dynamical mobility $\mu(\omega)$ constructed by the stochastic analytic continuation (SAC)~\cite{sm:Sandvik1998,sm:Shao2023,sm:Sandvik2016} and the maximum entropy (MaxEnt) methods~\cite{sm:SilverSiviaGubernatis1990,sm:KaufmannHeld2023}. For all representative temperatures and the Zeeman fields, we see that both methods yield qualitatively consistent results. In particular, at low temperatures, when the spectral peaks are narrow, the agreement is even quantitative. The consistency attests to the robustness of the analytic continuation.

\end{document}